\documentclass[runningheads]{llncs}
\usepackage{graphicx}
\usepackage{subfigure}
\usepackage{amsmath}
\usepackage{array}
\usepackage{multirow}
\usepackage{bbding}
\usepackage{graphicx}
\usepackage{multirow}
\usepackage{array}
\usepackage{makecell}

\usepackage[colorlinks,linkcolor=blue,citecolor=blue,urlcolor=blue]{hyperref}
 
\begin{document}

\title{FMFCC-A:  A  Challenging  Mandarin    Dataset for    Synthetic Speech Detection}

\titlerunning{FMFCC-A:  A  Challenging  Mandarin    Dataset}

\author
	{Zhenyu Zhang\inst{1,2}  \and
	Yewei Gu\inst{1,2} \and
	Xiaowei Yi\inst{1,2} \and
	Xianfeng Zhao$^{1,2(}$\Envelope$^{)}$}  
%
\authorrunning{Z. Zhang}
%
\institute{
State Key Laboratory of Information Security, Institute of Information Engineering, Chinese Academy of Sciences, Beijing 100093, China \\
\email{Zhaoxianfeng@iie.ac.cn \and
School of Cyber Security, University of Chinese Academy of Sciences, Beijing 100093, China
}}

\maketitle
\begin{abstract}
As increasing development of text-to-speech (TTS) and  voice conversion (VC) technologies, the detection of synthetic speech  has been suffered dramatically. 
In order to promote the  development of synthetic   speech detection model against  Mandarin TTS  and  VC technologies,  we have constructed a challenging Mandarin dataset   and organized the accompanying audio track of  the first fake media forensic challenge of China Society of Image and Graphics (FMFCC-A).
The FMFCC-A dataset is by far the largest  publicly-available Mandarin  dataset for  synthetic speech detection, which contains 40,000 synthesized  Mandarin  utterances that generated by 11  Mandarin TTS systems and two Mandarin VC systems, and 10,000 genuine Mandarin  utterances  collected  from 58 speakers.
The FMFCC-A dataset is divided into the training,  development and evaluation sets, which are used for the research of detection of synthesized  Mandarin speech under various  previously unknown  speech synthesis  systems  or audio post-processing operations.
In addition to describing  the construction of the FMFCC-A dataset, we provide a detailed analysis of two baseline methods and the top-performing submissions from the FMFCC-A, which illustrates the usefulness and challenge of FMFCC-A dataset. We hope that the FMFCC-A dataset can fill the gap of  lack of Mandarin datasets for synthetic speech detection.

\end{abstract}
\section{Introduction}
\label{sec:intro}

Synthetic speech attacks refer to a situation that an attacker makes use of text-to-speech (TTS) or voice conversion (VC) technologies to synthesize a speaker's voice to cheat the automatic speaker verification  (ASV) systems \cite{wu2015asvspoof}.
With the advancements in synthetic speech technologies, the state-of-the-art TTS and VC systems achieve such a high level of naturalness that even humans have difficulties to  distinguish  synthetic speech from genuine speech, which  imposes  a significant  threat to  the reliability of ASV systems \cite{todisco2019asvspoof}. 
Therefore, it is crucial to develop a synthetic speech detection model that can efficiently discriminate between genuine and synthetic speech \cite{kinnunen2017asvspoof}.

To protect ASV systems from  the increasing development of speech synthesis  technologies,  researchers from across the globe create fake speech detection challenges and release synthetic  speech datasets. 
Automatic speaker verification spoofing and countermeasures challenge,  called ASVspoof,  has been launched in 2015, 2017, 2019 and 2021, and four   speech datasets  (ASVspoof2015 dataset\cite{wu2015asvspoof}, ASVspoof2017 dataset \cite{kinnunen2017asvspoof}, ASVspoof2019 dataset \cite{todisco2019asvspoof},  ASVspoof2021 dataset \cite{yamagishi2021asvspoof}) have been released by the organizers. 
These ASVspoof datasets are   important milestones in the synthetic speech detection community, which  make the research community could study and propose methodologies to solve synthetic speech attacks to ASV systems. 
One of the most cited versions is ASVspoof2015 dataset \cite{wu2015asvspoof}, which contains not
only genuine utterances but also synthesized utterances generated by 10 different speech synthesis systems  (seven VC systems and three TTS systems), and in total more than 260,000 utterances are provided. ASVspoof2019 LA 
dataset \cite{todisco2019asvspoof} contains genuine utterances and synthesized utterances generated using 17 different TTS and VC systems in total 121,000 utterances. 
In addition, in order to make a speech dataset containing the utterances synthesized by the latest  speech synthesis technologies, Reimao et al. \cite{reimao2019dataset} released  a Fake or Real (FoR) dataset in 2019. 
However, it is important to note that the TTS and VC systems utilized in existing speech datasets are outdated compared to the current state-of-the-art speech synthesis systems, and all utterances are  in English. 
A new speech dataset containing the  synthesized speech by the latest speech synthesis technologies and  spoken in non-English is urgently needed to be proposed.

In order to facilitate the  development of synthetic  speech detection for synthesized  Mandarin  utterances,   we have constructed  a  challenging Mandarin dataset and organized the accompanying audio track of  the first fake media forensic challenge of China Society of Image and Graphics (FMFCC-A).
The FMFCC-A dataset is focused  on the latest Mandarin  speech synthesis technologies, where the synthesized  Mandarin speech is generated by not only open-source tools but also commercial  systems. 
The main contributions of this work are summarized as follows:
1)  We constructed the FMFCC-A dataset for the research of detection of synthesized  Mandarin speech and  divided it as the training,  development and evaluation sets, while four kinds of audio post-processing operations were conducted on valuation dataset. 
2) To demonstrate it is possible to use the FMFCC-A dataset to train deep learning based classifiers, we adopted two neural network based synthetic speech detection methods  as the baselines. 
3) We provided a detailed analysis of top-performing submissions from the FMFCC-A, which are focused upon  detection of Mandarin TTS and VC speech  under various previously unknown speech synthesis  systems or audio post-processing operations.
To  verify and reproduce the experiments  presented in the paper, all of our source codes and datasets are now available via GitHub: \url{https://github.com/Amforever/FMFCC-A}.

The remainder of this paper is organized as follows. In the next section, we formally introduced  the construction of the FMFCC-A dataset and two baseline methods. 
In Section 3, the evaluation metrics and   FMFCC-A schedule are described. 
In Section 4, we present a detailed analysis of two baseline methods and the top-performing submissions from the FMFCC-A. 
Finally, conclusions  and  potential future works are drawn in  the last section.

\section{Dataset and Baselines}
\label{sec:2MaterialsAndMethods}
 
The objective of this paper is  to construct a challenging Mandarin  dataset for  synthetic speech detection and  validate  the usefulness and challenge of the dataset.
In this section, we explain the construction of the FMFCC-A dataset  and two detection baselines  in detail.

\begin{table}[t]
	\caption{Summary of utterances in  the  FMFCC-A dataset, where G00 represents  the ID of genuine speech and A01 to A13  represent  the IDs of  13 speech synthesis systems.  }
	\label{DatasetDescript_1}
	\centering
	\small

			\scalebox{0.83}{\begin{tabular}{ p{1cm}<{\centering}  p{3.8cm}<{\centering}  c c c c }
			\hline
			
			\rule{0pt}{9pt} 	\multirow{2}{*}{ ID  }  & \multirow{2}{*}{Synthesis  systems}& \multicolumn{2}{c }{ Num. of speakers  }& \multirow{2}{*}{ Num. of utterances  } \\  
			\cline{3-4} 
			\rule{0pt}{9pt} 	& &Male&  Female& \\ 	\Xhline{1.2pt}
			
			G00	&	-	&	29	&	29	&	10,000	\\  \hline
			A01	&	 Alibaba  TTS	&	0	&	1	&	3,400	\\  \hline
			A02	&	 Biaobei TTS	&	5	&	5	&	3,400	\\  \hline
			A03	&	  Blackberry TTS	&	0	&	3	&	3,400	\\  \hline
			A04	&	 FastSpeech \cite{ren2020fastspeech} 	&	0	&	1	&	3,400	\\  \hline
			A05	&	Iflytek TTS 	&	2	&	2	&	3,400	\\  \hline
			A06	&	Ada-In VC \cite{chou2019one}	&	5	&	5	&	1,000	\\  \hline
			A07	&	IBM Waston TTS	&	1	&	2	&	4,000	\\  \hline
			A08	&	Lingyun  TTS	&	3	&	11	&	4000	\\  \hline
			A09	&	 Tacotron \cite{wang2017tacotron}	&	0	&	1	&	4,000	\\  \hline
			A10	&	Baidu  TTS	&	4	&	5	&	3,000	\\  \hline
			A11	&	Sibichi TTS	&	3	&	3	&	3,000	\\  \hline
			A12	&	  GAN-TTS \cite{binkowski2019high}	&	1	&	0	&	3,000	\\  \hline
			A13	&	 Medium  VC  \cite{gu2021mediumvc}	&	5	&	5	&	1,000	\\  \hline
			\multicolumn{2}{ c }{\textbf{Total}}	&	58	&	73	&	50,000	\\  \hline
			
	\end{tabular}}
\end{table} 

\subsection{FMFCC-A Dataset and Partitions}
\label{CSSDDatabase}
The FMFCC-A dataset contains not only genuine utterances but also a mountain of synthesized utterances, which are spoken in Mandarin.
The total number of utterances of the FMFCC-A dataset is 50,000, which includes 10,000 genuine utterances and 40,000 synthesized utterances. 
The genuine utterances, i.e. speech recordings from humans, were randomly collected from 58 speakers,  which covers a good variety of speaker ages and genders. 
The synthesized utterances  were generated from state-of-the-art  Mandarin speech synthesis systems. 
Specifically, the synthesized utterances of the FMFCC-A dataset  were generated according to 11  Mandarin TTS systems,  i.e., Alibaba  TTS, Biaobei TTS, Blackberry TTS, FastSpeech \cite{ren2020fastspeech}, Iflytek TTS,  IBM Waston TTS,  Lingyun TTS, Tacotron \cite{wang2017tacotron}, Baidu TTS, Sibichi TTS and GAN-TTS \cite{binkowski2019high},  and two Mandarin VC systems, i.e., Ada-In VC \cite{chou2019one} and Medium  VC  \cite{gu2021mediumvc}. The 13 Mandarin speech synthesis systems are noted  as A01 to A13 in this paper (as in Table \ref{DatasetDescript_1}), where A04 \cite{ren2020fastspeech}, A06 \cite{chou2019one}, A09 \cite{wang2017tacotron}, A12 \cite{binkowski2019high} and A13 \cite{Lin2021S2VCAF} are open-source    speech synthesis  tools  and A01, A02, A03, A05, A07, A08, A10 and A11 are  commercial   speech synthesis systems, 
For five open-source synthesis  tools, we generated synthesized Mandarin utterances by implementing  their source codes, while 
we sent  API requests for commercial  speech synthesis systems to get  synthesized Mandarin utterances.
For each utterance in the FMFCC-A dataset, the duration is randomly set in the range between two seconds and 10 seconds, the sampling rate of 16 kHz, 16-bit quantization and is stored in mono WAV format. More details about the utterances in the FMFCC-A dataset are shown in Table \ref{DatasetDescript_1}.

As with  most  researches of synthetic speech detection, 
the FMFCC-A dataset is partitioned into three disjoint datasets, namely training, development and evaluation sets.   More details about the  utterances and the used  speech synthesis systems in training, development and evaluation datasets are shown in Table \ref{DatabasePartitions}.
The training dataset includes 6,000 synthesized utterances and  4,000 genuine utterances. As illustrated in Table  \ref{DatabasePartitions}, each synthesized utterance in training dataset is generated  by one of the five  speech synthesis  systems (A01 – A05).
The development dataset includes 17,000 synthesized utterances and  3,000 genuine utterances. 
In order to generalize the performance of synthetic   speech detection models  against previously unknown  speech synthesis  systems, we also produced synthetic speech with five additional speech synthesis systems (A06 – A09).   

\begin{table}[!ht]
	\renewcommand{\arraystretch}{1}
	\caption{The number of  utterances and corresponding speech synthesis systems  in the training (Train), development (Dev) and evaluation (Eval) sets of  FMFCC-A dataset.  } 
	\small
	\label{DatabasePartitions}
	\centering
	\scalebox{0.83}{\begin{tabular}{p{1cm}<{\centering}   c  p{0.8cm}<{\centering} p{0.8cm}<{\centering} p{0.8cm}<{\centering} p{0.8cm}<{\centering}p{0.8cm}<{\centering} p{0.8cm}<{\centering}p{0.8cm}<{\centering} p{0.8cm}<{\centering}p{0.8cm}<{\centering} p{0.8cm}<{\centering}p{0.8cm}<{\centering} p{0.8cm}<{\centering}p{0.8cm}<{\centering} p{0.9cm}<{\centering} }
			\hline
			Subset	&	G00	&	A01	&	A02	&	A03	&	A04	&	A05	&	A06	&	A07	&	A08	&	A09	&	A10	&	A11	&	A12	&	A13	&	 Total		\\  \Xhline{1.2pt}
			Train	&	4,000	&	1,200	&	1,200	&	1,200	&	1,200	&	1,200	&	0	&	0	&	0	&	0	&	0	&	0	&	0	&	0	&	10,000		\\  \hline
			Dev 	&	3,000	&	1,600	&	1,600	&	1,600	&	1,600	&	1,600	&	900	&	2,700	&	2,700	&	2,700	&	0	&	0	&	0	&	0	&	20,000		\\  \hline
			Eval 	&	3,000	&	600	&	600	&	600	&	600	&	600	&	100	&	1,300	&	1,300	&	1,300	&	3,000	&	3,000	&	3,000	&	1,000	&	20,000		\\  \hline

	\end{tabular}}
\end{table}

\begin{figure*}[t]
	\centering
	\centerline{\includegraphics[width=0.88\textwidth]{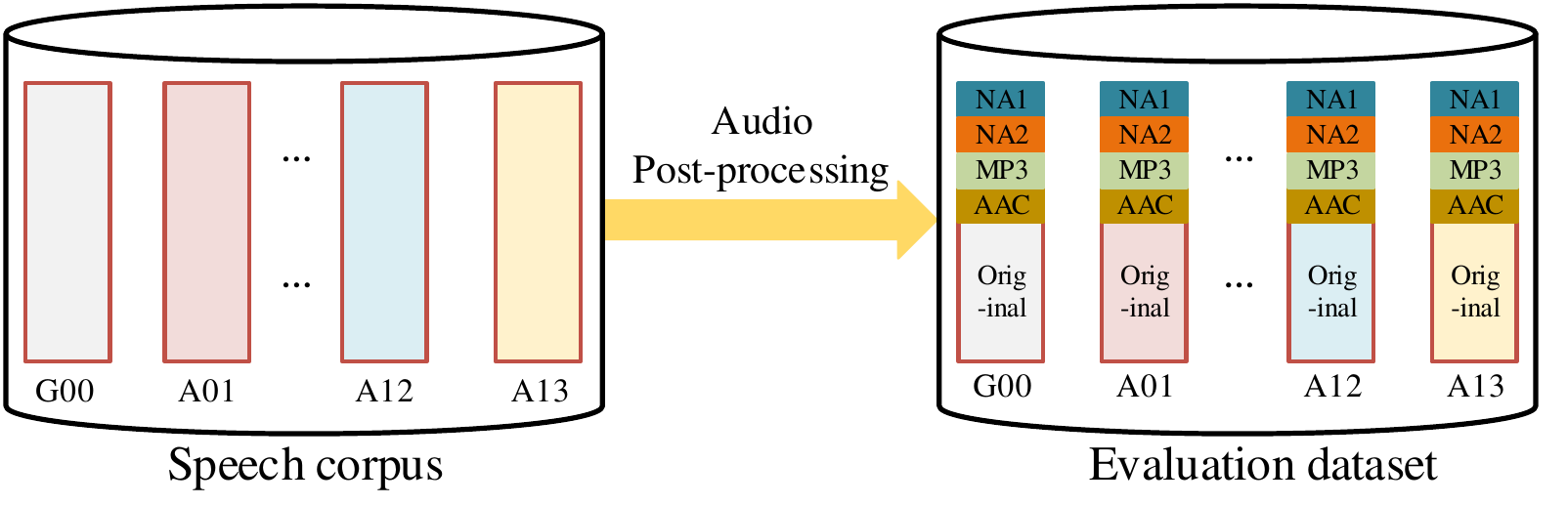}}
	\caption{The flowchart of performing four audio post-processing operations on the evaluation set of the FMFCC-A dataset.}
	\label{fig:EvalDatasetPostProcessing}
\end{figure*}

The  evaluation set of  FMFCC-A dataset  is designed to evaluate and analyze the synthetic speech detection models under previously unknown speech synthesis systems and audio post-processing operations, so the construction of evaluation dataset is more complicated than training and development sets (as shown in Fig. \ref{fig:EvalDatasetPostProcessing}), 
The evaluation dataset is comprised of  17,000 synthesized utterances and 3,000 genuine  utterances.
On the one hand, to evaluate the synthetic   speech detection models against previously unknown speech synthesis systems, the synthesized utterances are generated according to more diverse speech synthesis systems which include the same nine  speech synthesis systems (A01 - A09) used   in  the development dataset and  additional four  speech synthesis systems (A10 - A13).
On the other  hand, in order to evaluate the performance of  synthetic speech detection models under various audio post-processing operations, 50 percent of the evaluation dataset are randomly selected to undergone compression-decompression operation or  additive Gaussian noise operation.
Specifically, we  selected 12.5 percent of    the evaluation dataset  and transformed them (WAV files) into MPEG-1 audio layer 3 (MP3) with 96 kilobits per second (Kbps)  and converted MP3 files back to WAV using FFmpeg software\footnote[1]{http://ffmpeg.org/}. In the same way, advanced audio coding (AAC) compression of 64 Kbps and decompression  were conducted on another 12.5 percent of the evaluation dataset.
For random Gaussian noise  addition, we   selected 12.5 percent of   the evaluation dataset and process the speech with 0.01  random Gaussian noise addition, and  process another   12.5 percent of   the evaluation dataset with 0.002 random Gaussian noise addition. 
Finally, 50 percent of the evaluation dataset are  undergone a compression-decompression operations or  additive Gaussian noise operations and the remaining 50 percent of the evaluation dataset are  original without any audio post-processing operations.

\begin{figure*}[t]
	\centering
	\centerline{\includegraphics[width=0.96\textwidth]{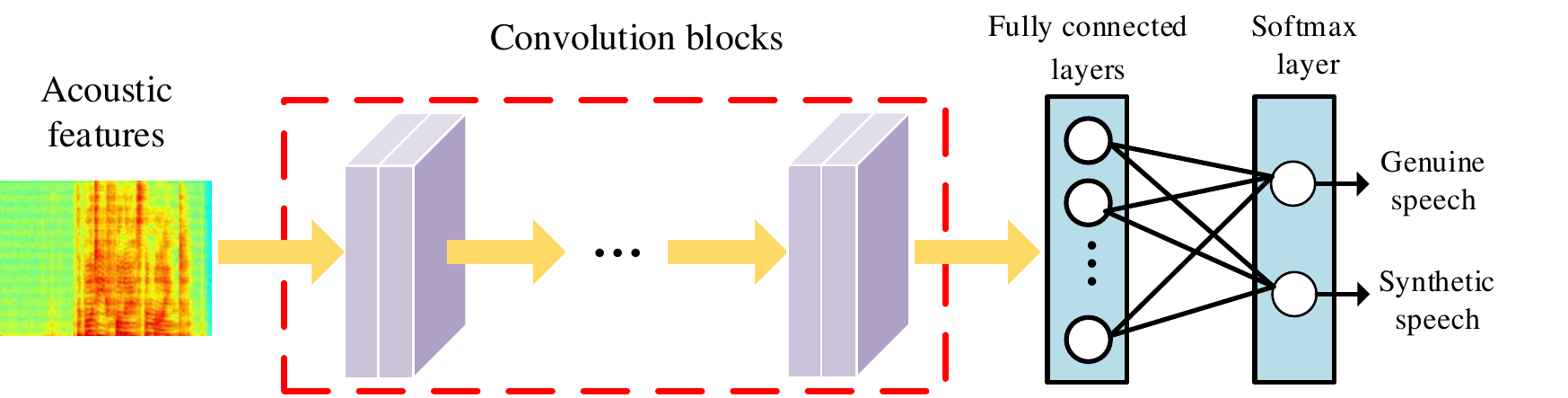}}
	\caption{The flowchart of the neural networks based synthetic speech detection methods. }\label{fig:CNNFlowDigram}
\end{figure*} 

\subsection{Baseline Methods}
 
The whole flowchart of the neural networks based  synthetic   speech detection models   is illustrated  in Fig. \ref{fig:CNNFlowDigram}. 
Firstly, acoustic features are extracted from the speech signal. Then, several different designed convolution blocks are stacked to extract contextual representations of the input acoustic feature. Finally, the fully connected layers map the deep features on the label space of speech and the probability of being synthetic speech is calculated by a softmax layer. 
In recent years, many  acoustic features  have been proposed for synthetic speech detection, while Mel-frequency cepstral coefficient (MFCC) \cite{patel2015combining} and constant-Q cepstral coefficient (CQCC) \cite{todisco2016new} are two of the most used acoustic features.
In addition,   light convolutional neural network (LCNN) \cite{bialobrzeski2019robust}  and residual convolutional neural network (ResNet) \cite{alzantot2019deep} are  the popular neural network architectures used as deep representation learning   and classification for synthetic speech detection.  
Nowadays, several end-to-end synthetic speech detection models based on neural networks with mere the raw speech waveform have been proposed \cite{tak2021end,hua2021towards}, which could achieve satisfactory performance.

In order to demonstrate it is possible to train deep learning based classifiers using the FMFCC-A dataset, we adopted two neural network  based synthetic speech detection methods  as the baselines.  
The first baseline method (B01) consists of 34 layers ResNet and CQCC feature \cite{todisco2016new}. The parameters of extraction of the CQCC feature are similar to those in  \cite{todisco2019asvspoof}. The second baseline  method (B02) adopted  Res-TSSDNet architecture  \cite{hua2021towards} and the raw speech waveform, which is an end-to-end synthetic speech detection method. 
Two baseline  methods are implemented  in PyTorch  1.1.0\footnote[2]{https://pytorch.org/}, which are trained on the training set of FMFCC-A dataset, and tested on the development and evaluation sets respectively.

\section{Evaluation Metrics and FMFCC-A Schedule}

\subsection{Evaluation Metrics}
Two  `threshold-free' evaluation metrics are used to estimate performances of  synthetic speech detection methods on  the FMFCC-A dataset, which are  
Log-loss \cite{vovk2015fundamental} and  equal error rate (EER) \cite{kinnunen2017asvspoof}.
The Log-loss \cite{vovk2015fundamental} is one of the major metrics to assess the performance of a classification problem. It is indicative of how close the prediction probability is to the corresponding true value (0 or 1 in the case of binary classification). The more predicted probability diverges from the actual value, the higher is the log-loss value. For binary classification, Log-loss can be expressed as follows:
\begin{equation}
\label{LOSSlogFormula}
\textrm{Log-loss}_{i}=-[y_{i} {\rm ln}p_{i}+(1-y_{i}){\rm ln}(1-p_{i})],
\end{equation}
where $i$ is the index of utterances and it corresponding label is $y_{i}$; $p_{i}$ is the prediction probability and ln refers to the natural logarithm of a number. 
To avoid the case of ln(0),  ln(max(p), $\epsilon$) is computed where $\epsilon$ is a positive real number closing to zero.  The lower Log-loss indicates that the detection models have performed better. 

The EER  is a commonly used metric for the  datasets that  are heavily unbalanced for two-class classification  (one genuine utterance versus many   synthesized utterances) \cite{kinnunen2017asvspoof}.
Suppose $P_{\textrm{fa}}(\theta)$ and $P_{\textrm{fr}}(\theta)$ stand for the false acceptance rate  and  false rejection rate  respectively, which are defined as :

\begin{equation}
P_{\textrm{fa}}(\theta) = \frac{\mathrm{Num\{ synthesized\ utterances \ with\ score > \theta \}}}{ \mathrm{Num\{total\ synthesized\ utterances\}}},
\end{equation}

\begin{equation}
P_{\textrm{fr}}(\theta) = \frac{\mathrm{Num\{genuine\ utterances \ with\ score \leq \theta \}}}{ \mathrm{Num\{total\ genuine\ utterances\}}},
\end{equation}
where   $\mathrm{Num}{\{ \cdot \}}$  denotes the number of utterances in the set. The EER corresponds to the threshold $\theta_{\textrm{EER}}$ at which the   false acceptance rate  equals to  false rejection rate i.e. $\mathrm{EER} = P_{\textrm{fa}}(\theta_{\textrm{EER}}) = P_{\textrm{fr}}(\theta_{\textrm{EER}})$.
In this paper, the EER is estimated using the Bosaris  toolkit\footnote[3]{https://sites.google.com/site/bosaristoolkit/}.

\subsection{FMFCC-A Schedule}
FMFCC-A  focuses on the research of  detection of synthesized  Mandarin speech, which is run in two phases: the preliminary phase and the main evaluation phase.   
Participants can use training and development sets of the FMFCC-A  dataset to construct and optimize  their synthetic speech detection models, while the use of  other external speech datasets is also allowed. Participants should assign each utterance with a real-valued and finite score range in [0, 1] which reflected the relative probability that the test utterance was genuine. The Log-loss \cite{vovk2015fundamental} was adopted as the metric for evaluating the performance of submissions, and $\epsilon$ in Formula (\ref{LOSSlogFormula}) is set to 10e$^{-9}$. 

During the preliminary phase,  the training and development sets of FMFCC-A  dataset were available for each participant, and a standard protocol file comprising a list of filenames and corresponding labels of utterances from the training dataset was provided.
However, the label information of speech from the development dataset  was not provided.  Each participant team might make up to 3 submissions per day and the corresponding validation  performance of the submission  would be provided  by the organizer.  
The focus of the preliminary phase was upon the training  and optimizing synthetic speech detection methods that were robust to previously unknown speech synthesis systems.
During the main evaluation phase, participants were allowed only a single submission for which the result would be determined by the utterances from the evaluation dataset. Given the relevance to realistic  scenarios, the focus of the main evaluation phase was upon  evaluating generalization of synthetic speech detection methods   to previously unknown speech synthesis systems and audio post-processing operations.

\section{FMFCC-A Results and Discussion}

\subsection{Preliminary  Phase of FMFCC-A} 
In the preliminary phase of the FMFCC-A, a total of 48 valid submissions were received. 
Since the aim of the FMFCC-A is the scientific analysis of different Mandarin synthetic speech detection methods, we assign anonymized  team IDs (T01 to T48) to participators, which are corresponding to the order ranking in the preliminary phase.
Table  \ref{PreliminaryProgressResultTable} shows results in terms of the Log-loss and EERs for all submissions as well as two baseline methods (B01 and B02), pooled over all speech synthesis systems in the development dataset. 

\begin{table}[t]
	\renewcommand{\arraystretch}{1}
	\caption{Log-loss and EER  (\%) scores of all submissions and two baseline methods as measured on the development set of the FMFCC-A dataset.}
	\small
	\label{PreliminaryProgressResultTable}
	\centering
	\scalebox{0.9}{\begin{tabular}{  p{0.6cm}   l   p{2cm}<{\centering}    p{1.2cm} | p{0.8cm}<{\centering}    p{1.2cm}   p{2cm}     p{1.2cm}   }
			\hline
			\#  & ID & Log-loss & EER&\# & ID & Log-loss &  EER\\
			\Xhline{1.2pt}
			1	&	T01	&	0.000083 	&	0.00 	&	26	&	T26	&	0.471997 	&	9.40 	\\  \hline
			2	&	T02	&	0.000223 	&	0.00 	&	27	&	T27	&	0.536461 	&	73.40 	\\  \hline
			3	&	T03	&	0.002369 	&	0.07 	&	28	&	\textbf{B1}	&	\textbf{0.539770}	&	\textbf{7.27}	\\  \hline
			4	&	T04	&	0.026720 	&	0.40 	&	29	&	T28	&	0.550739 	&	7.13 	\\  \hline
			5	&	T05	&	0.055830 	&	2.83 	&	30	&	T29	&	0.551340 	&	4.73 	\\  \hline
			6	&	T06	&	0.085988 	&	1.94 	&	31	&	T30	&	0.594707 	&	3.70 	\\  \hline
			7	&	T07	&	0.086400 	&	1.44 	&	32	&	T31	&	0.608583 	&	85.64 	\\  \hline
			8	&	T08	&	0.088507 	&	2.83 	&	33	&	T32	&	0.631574 	&	17.83 	\\  \hline
			9	&	T09	&	0.088626 	&	2.30 	&	34	&	T33	&	0.671077 	&	8.13 	\\  \hline
			10	&	T10	&	0.157802 	&	7.93 	&	35	&	T34	&	0.678155 	&	8.33 	\\  \hline
			11	&	T11	&	0.171027 	&	11.03 	&	36	&	T35	&	0.686454 	&	98.63 	\\  \hline
			12	&	T12	&	0.171333 	&	11.03 	&	37	&	T36	&	0.695494 	&	99.94 	\\  \hline
			13	&	T13	&	0.171480 	&	11.23 	&	38	&	T37	&	0.699639 	&	99.97 	\\  \hline
			14	&	T14	&	0.174231 	&	7.33 	&	39	&	\textbf{B2}	&	\textbf{0.815244}	&	\textbf{8.26}	\\  \hline
			15	&	T15	&	0.175598 	&	4.27 	&	40	&	T38	&	0.842023 	&	33.27 	\\  \hline
			16	&	T16	&	0.175821 	&	12.29 	&	41	&	T39	&	0.933437 	&	20.77 	\\  \hline
			17	&	T17	&	0.196450 	&	7.33 	&	42	&	T40	&	1.549387 	&	42.54 	\\  \hline
			18	&	T18	&	0.204748 	&	7.43 	&	43	&	T41	&	1.905160 	&	22.03 	\\  \hline
			19	&	T19	&	0.213509 	&	7.30 	&	44	&	T42	&	2.247894 	&	14.17 	\\  \hline
			20	&	T20	&	0.258818 	&	12.63 	&	45	&	T43	&	2.713916 	&	50.77 	\\  \hline
			21	&	T21	&	0.348626 	&	12.60 	&	46	&	T44	&	2.935695 	&	50.33 	\\  \hline
			22	&	T22	&	0.351485 	&	14.17 	&	47	&	T45	&	3.614879 	&	56.30 	\\  \hline
			23	&	T23	&	0.405014 	&	3.93 	&	48	&	T46	&	4.128621 	&	99.50 	\\  \hline
			24	&	T24	&	0.421419 	&	30.93 	&	49	&	T47	&	6.565136 	&	27.46 	\\  \hline
			25	&	T25	&	0.445320 	&	26.73 	&	50	&	T48	&	23.486387 	&	100.00 	\\  \hline

	\end{tabular}}
\end{table} 

As we can see from Table  \ref{PreliminaryProgressResultTable},  
27 of 48 participating teams submitted  results that outperformed the B01  in terms of the Log-loss, and 13 participating teams'  results achieved  lower EERs than   that of B01. 
The top two performing submissions, T01 and T02, achieve   Log-loss of 0.000083 and 0.000223, and both EERs  are  equal to  0, which are performing results.
However, it observes that monotonic increases in the Log-loss that are not always mirrored by monotonic increases in the EER.
The  participating teams T15, T23, T30, and T31 deliver low EERs, however, they have high Log-loss values which lead them to a low ranking in the preliminary phase of FMFCC-A.
There is a great spread both in Log-loss and EERs, however, the performance  of the top nine  performing systems is much narrower.  Top nine  performing submissions  achieve Log-loss below 0.1 and EERs below 3\%  even when   almost  half of  the development dataset are generated by  previously unknown   speech synthesis systems, which deliver a substantial improvement over other submissions. 
After communications with participants, we knew that some of   the participants used the idea of pseudo-label learning methods. 
They uploaded result files, which contained scores of all  utterances on the development dataset, to get the feedback from the organizers and sent the  feedback to  their neural networks to make the neural networks  more effectively learn for difficult samples.
The performance of submission on preliminary phase of FMFCC-A  demonstrates that the training set of  FMFCC-A dataset could be used to train deep learning based classifiers, while the development set of  FMFCC-A dataset could be further  optimized the parameters of synthetic speech detection models.

\begin{table}[t]
	\renewcommand{\arraystretch}{1}
	\caption{Log-loss and EER (\%) scores of  top 10  performing  submissions  and two baseline methods as measured on  the evaluation set of the FMFCC-A dataset.}
	\small
	\label{Final_1ProgessPool}
	\centering
	\scalebox{0.9}{\begin{tabular}{  p{0.6cm}<{\centering}   c   p{2cm}<{\centering}    p{1.2cm}<{\centering} | p{0.6cm}<{\centering}   c   p{2cm}<{\centering}    p{1.2cm}<{\centering}  }
			\hline
			\#   & ID  &   Log-loss  &   EER  & \#  &  ID  &   Log-loss  &   EER \\
			\Xhline{1.2pt}
			1	&	T06	&	0.351378 	&	13.59 	&	7	&	T05	&	1.070571 	&	22.86 		\\  \hline
			2	&	T01	&	0.367338 	&	9.50 	&	8	&	T09	&	1.504904 	&	27.70 		\\  \hline
			3	&	T03	&	0.417512 	&	12.57 	&	9	&	T11	&	1.739203 	&	23.83 		\\  \hline
			4	&	T04	&	0.470891 	&	12.00 	&	10	&	T10	&	1.952625 	&	24.93 		\\  \hline
			5	&	T02	&	0.593899 	&	17.23 	&	11	&	\textbf{B01}	&	\textbf{2.133188} 	&	\textbf{28.07} 		\\  \hline
			6	&	T07	&	0.694353 	&	23.00 	&	12	&\textbf{B02}	&	\textbf{2.286764} 	&	\textbf{33.37} 		\\  \hline
			
	\end{tabular}}
\end{table}

\subsection{Main Evaluation Phase of FMFCC-A}

In the main evaluation phase, the top 10 participants in the preliminary phase were required to run their submissions on the evaluation set of the FMFCC-A dataset. Note that the eighth-place team (T08) in the preliminary phase, unfortunately, did not run their submission  despite repeated warnings from the organizers. Therefore, we excluded them in the  main evaluation phase and asked the eleventh-place participating team (T11) in the preliminary phase   to participate in the main evaluation phase.
The Log-loss and EER results of 12 synthetic speech detection methods on the main evaluation phase are illustrated in Table \ref{Final_1ProgessPool}.

As can be seen in Table \ref{Final_1ProgessPool}, the T06, which ranks sixth place on the preliminary  phase, achieves the first place in the main evaluation phase with a Log-loss of 0.351378 and EER of 13.59\%.  The T01 that achieves the best performance in  the preliminary  phase ranks second place in the main evaluation phase  according to the  Log-loss. However, T01 delivers the lowest EER of 9.50\% among all submissions. 
In addition,  all top ten performing submissions achieve better performance than two baseline methods in the main evaluation phase.
The Log-loss of all synthetic speech detection models on the main evaluation phase are ranged between 0.351378 and 2.286764, and EER are ranged between 9.50 and 33.37, which are increased a lot compared to those on the preliminary  phase.   
This observation indicates that the evaluation set of the FMFCC-A dataset has dramatically decreased the performances of submissions and baselines,  which illustrates the challenge of the FMFCC-A dataset for  synthetic speech detection methods. 
To evaluate the performance of submissions  against    previously unknown speech synthesis systems and audio post-processing operations, we give a detailed description  in the following.

\begin{table}[!ht]
	\renewcommand{\arraystretch}{1}
	\caption{EERs (\%)  of  detection methods as measured on the evaluation set of the FMFCC-A dataset  subjected to each  speech synthesis system and pooled performances.}
	\small
	\label{FinalProgess_2SingleAttackPool}
	\centering
	\scalebox{0.836}{\begin{tabular}{p{0.58cm}<{\centering} p{0.76cm}<{\centering} p{0.76cm}<{\centering} p{0.76cm}<{\centering} p{0.76cm}<{\centering}p{0.76cm}<{\centering} p{0.76cm}<{\centering}p{0.76cm}<{\centering} p{0.76cm}<{\centering}p{0.76cm}<{\centering} p{0.76cm}<{\centering}p{0.76cm}<{\centering} p{0.76cm}<{\centering}p{0.76cm}<{\centering}p{0.76cm}<{\centering} p{0.76cm}<{\centering}p{0.76cm}<{\centering}}
			\hline
			ID	&	A01	&	A02	&	A03	&	A04	&	A05	&	Pool1	&	A06	&	A07	&	A08	&	A09	&	Pool2	&	A10	&	A11	&	A12	&	A13	&	Pool3	\\ [1.8pt]  	\Xhline{1.2pt}
			T06	&	2.32 	&	2.33 	&	1.33 	&	2.15 	&	3.00 	&	2.23 	&	19.08 	&	6.61 	&	11.45 	&	2.84 	&	8.06 	&	9.57 	&	3.10 	&	24.30 	&	39.10 	&	19.20 	\\ [1.8pt]  \hline			
			T01	&	1.53 	&	5.67 	&	0.30 	&	2.48 	&	2.00 	&	2.67 	&	9.77 	&	1.09 	&	2.61 	&	1.09 	&	1.90 	&	5.43 	&	4.23 	&	12.83 	&	53.70 	&	12.70 	\\ [1.8pt]  \hline			
			T03	&	1.52 	&	6.02 	&	1.82 	&	2.83 	&	5.67 	&	3.80 	&	12.00 	&	2.32 	&	10.98 	&	7.70 	&	6.96 	&	6.30 	&	4.27 	&	20.83 	&	42.28 	&	16.93 	\\ [1.8pt]  \hline			
			T04	&	1.67 	&	2.52 	&	0.70 	&	3.65 	&	5.50 	&	3.63 	&	20.78 	&	10.07 	&	11.00 	&	7.16 	&	9.90 	&	7.37 	&	2.83 	&	17.03 	&	59.40 	&	16.00 	\\ [1.8pt]  \hline			
			T02	&	0.02 	&	2.67 	&	1.83 	&	0.67 	&	0.67 	&	1.30 	&	7.00 	&	2.62 	&	3.23 	&	0.54 	&	2.46 	&	10.63 	&	1.57 	&	27.30 	&	55.20 	&	22.83 	\\ [1.8pt]  \hline			
			T07	&	3.35 	&	9.13 	&	5.53 	&	2.82 	&	4.32 	&	5.47 	&	30.00 	&	12.61 	&	8.70 	&	5.07 	&	10.53 	&	9.10 	&	9.17 	&	47.47 	&	61.80 	&	33.13 	\\ [1.8pt]  \hline			
			T05	&	8.02 	&	10.30 	&	15.63 	&	9.50 	&	12.00 	&	11.60 	&	22.00 	&	18.16 	&	17.84 	&	16.39 	&	17.53 	&	17.27 	&	16.80 	&	37.67 	&	56.90 	&	29.20 	\\ [1.8pt]  \hline			
			T09	&	16.97 	&	15.35 	&	13.50 	&	18.82 	&	21.17 	&	16.90 	&	42.00 	&	19.32 	&	27.14 	&	20.75 	&	22.17 	&	20.87 	&	15.43 	&	44.03 	&	51.78 	&	32.87 	\\ [1.8pt]  \hline			
			T11	&	23.80 	&	23.80 	&	23.80 	&	23.80 	&	23.80 	&	23.77 	&	42.03 	&	23.77 	&	23.77 	&	23.77 	&	23.77 	&	23.77 	&	28.23 	&	31.63 	&	24.30 	&	28.14 	\\ [1.8pt]  \hline			
			T10	&	3.33 	&	4.33 	&	1.18 	&	6.33 	&	5.00 	&	4.37 	&	44.85 	&	15.70 	&	15.23 	&	10.45 	&	14.70 	&	7.43 	&	5.17 	&	51.10 	&	55.38 	&	35.11 	\\ [1.8pt]  \hline			
			B01	&	9.50 	&	17.83 	&	3.00 	&	4.63 	&	14.80 	&	10.33 	&	55.12 	&	21.07 	&	21.61 	&	5.54 	&	18.60 	&	32.73 	&	32.47 	&	34.60 	&	62.92 	&	36.40 	\\ [1.8pt]  \hline			
			B02	&	18.50 	&	27.17 	&	16.83 	&	19.67 	&	20.98 	&	20.67 	&	47.13 	&	38.39 	&	29.38 	&	20.39 	&	31.27 	&	31.40 	&	30.40 	&	47.67 	&	40.92 	&	36.87 	\\ [1.8pt]  \hline			
			Avg.	&	7.54 	&	10.59 	&	7.12 	&	8.11 	&	9.91 	&	8.89 	&	29.31 	&	14.31 	&	15.24 	&	10.14 	&	13.99 	&	15.16 	&	12.81 	&	33.04 	&	50.31 	&	26.61 	\\ [1.8pt]  \hline

	\end{tabular}}
\end{table}

\subsection{Performance on Unknown Speech Synthesis Systems}
While the participants' rankings of FMFCC-A  are based on pooled performance (the synthetic speech from all speech synthesis systems in the dataset are considered for evaluation), it is also meaningful to present  results when decomposed by each  speech synthesis system. 
As described in Section \ref{CSSDDatabase}, we classify the evaluation set of the FMFCC-A dataset into three parts according to whether the speech synthesis systems are used in training and development datasets. The known  speech synthesis systems (KSSS) set includes A01 to A05 which are presented in both of the training and development datasets.  The  first unknown speech synthesis systems (USSS1) set includes A06 to A09 which are not presented in  the training dataset but used in the development dataset.
The second unknown  speech synthesis systems  (USSS2) set  includes A10 to A13 which are neither used in training dataset nor development dataset. The EER results  when decomposed separately for each of the speech synthesis systems are presented   in  Table \ref{FinalProgess_2SingleAttackPool},  while the pooled detection results for KSSS, USSS1, and USSS2  (named as Pool1, Pool2 and Pool3 respectively) are also presented.

As shown in Table \ref{FinalProgess_2SingleAttackPool},  each line represents the EERs of one  detection method against various speech synthesis systems. 
The speech synthesis systems A06, A12 and A13 seriously degrade the performances of detection methods, which showed being challenging to be detected.
In fact, the A06 and A13 are Mandarin VC systems  and A12 is a generative adversarial network (GAN) based Mandarin TTS system, which are fundamentally different from other Mandarin TTS systems.
Therefore, the  detection methods that   trained on the  training dataset achieve   poor performance for  A06, A12 and A13. 
Although  A07, A08, A10, and A11 are also  threat detection methods, they are easier to detect than A06, A12 and A13.
One reason may be that A07, A08, A10, and A11 are pipeline TTS techniques that are similar mechanism with TTS techniques of A01 to A05.
In addition,   EERs of Pool1, Pool2 and Pool3 are 	8.89 \%, 	13.99 \% and 26.61 \% respectively, which means  that the threat of previously unknown speech synthesis systems to  detection methods  are obviously  compared to the known speech synthesis systems.

\begin{figure*}[t]
	\centering
	\centerline{\includegraphics[width=0.92\textwidth]{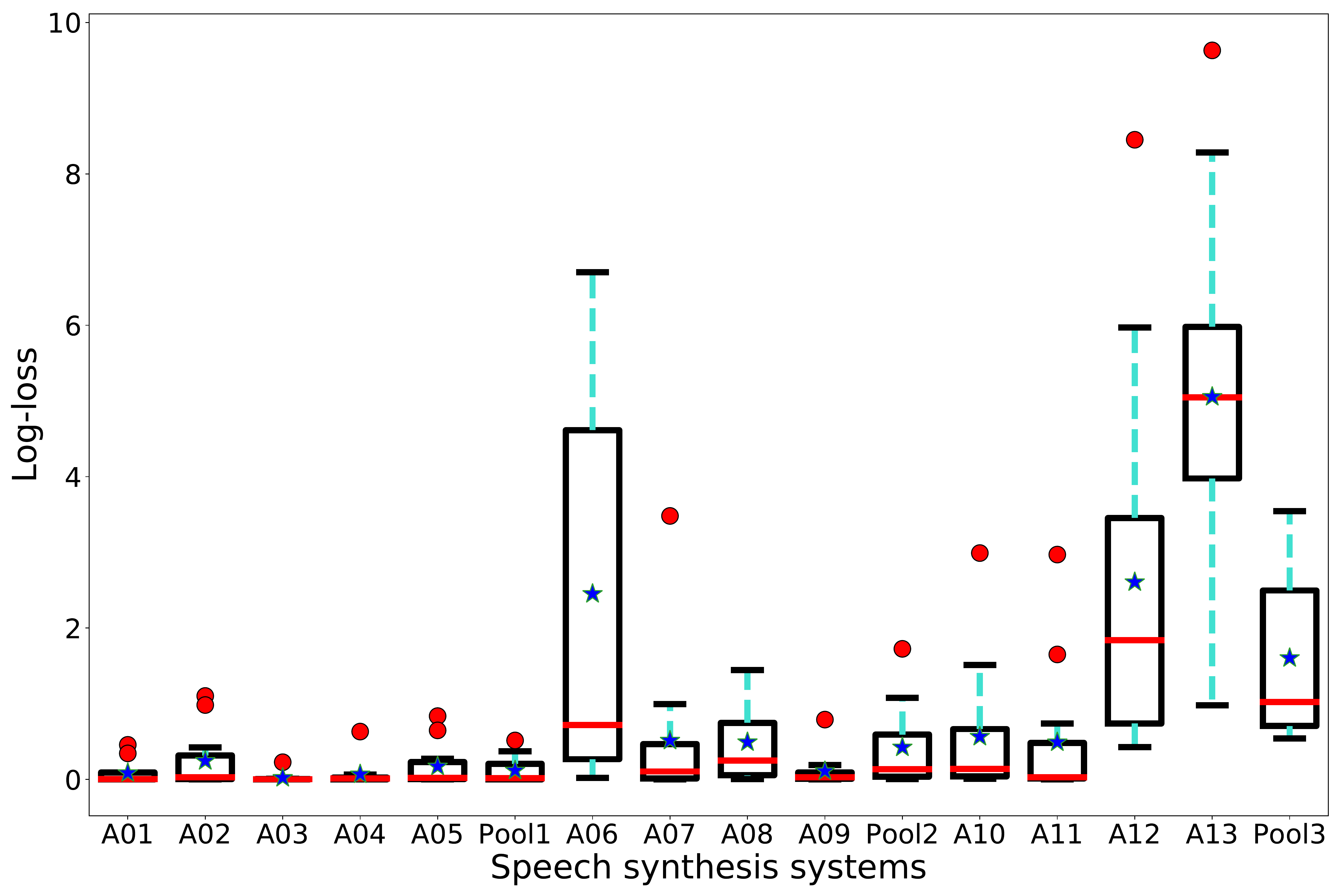}}
	\caption{A boxplot of Log-loss of  detection methods as measured on the evaluation set of the FMFCC-A dataset   subjected to each  speech synthesis system and pooled performances.}  
	\label{fig:CNNDigram}
\end{figure*}

In order to show the variation of Log-loss results on the evaluation dataset for  synthetic speech detection methods,  we illustrate  boxplots Log-loss results in Fig. \ref{fig:CNNDigram}. The similar trends can be observed in  the boxplots of Log-loss.
There is a greater spread in Log-loss for attacks A06, A12 and A13, which shown they are a greater threat to detection methods.
The difference  between performances of detection methods against A01 to A05   is much narrower. 
Through the analysis of  detection methods for known attacks (KSSS) and unknown attacks (USSS1 and USSS2), it indicates the importance of improving the detection models against  previously unknown speech synthesis systems.

\subsection{Performance on Audio Post-processing Operations} 

Considering the relevance to realistic scenarios, it is also very important for the detection models to detecting synthesized  Mandarin utterances that have undergone various audio post-processing operations. In this section, we analyze the results of detection models for the synthesized  Mandarin utterances undergone each audio post-processing operation on the evaluation set of the FMFCC-A dataset.
As described in Section \ref{CSSDDatabase}, we classified the evaluation dataset into five parts according to whether utterances have undergone post-processing operations. Firstly, the 50 percent of  the evaluation dataset were not undergone any audio post-processing operations  as original set (Original). Secondly, the 12.5 percent of  the evaluation dataset were processed by MP3 compression-decompression with 96 Kbps (CD-MP3). Thirdly, the 12.5 percent of  the evaluation dataset were processed by AAC compression-decompression  with 64 Kbps (CD-AAC). Next,  the 12.5 percent of  the evaluation dataset were added  with 0.01 random Gaussian noise  (NA-01). Finally, the remaining 12.5 percent of  the   evaluation dataset were added  with 0.002 random Gaussian noise  (NA-002). 
Table \ref{FinalProgess_4PostProcessing} shows the EERs of detection models for detecting original and audio post-processing utterances in the evaluation dataset.

\begin{table}[t]
	\renewcommand{\arraystretch}{1}
	\caption{EERs (\%) of    detection methods as measured on the evaluation set of the FMFCC-A dataset  under  original and audio post-processing utterances.}
	\small
	\label{FinalProgess_4PostProcessing}
	\centering
	\begin{tabular}{  p{1.2cm}<{\centering}   c c c  c c  }
		\hline
		ID &   Original &  NA-01 &  NA-002 &  CD-MP3  &  CD-AAC \\
		\Xhline{1.2pt}
		T06	&	9.54 	&	23.67 	&	20.54 	&	12.26 	&	11.16 	\\  \hline
		T01	&	6.92 	&	11.20 	&	10.97 	&	7.78 	&	7.20 	\\  \hline
		T03	&	7.45 	&	16.52 	&	12.52 	&	7.48 	&	8.25 	\\  \hline
		T04	&	8.99 	&	18.67 	&	14.12 	&	10.13 	&	9.86 	\\  \hline
		T02	&	15.94 	&	21.67 	&	21.12 	&	16.03 	&	18.13 	\\  \hline
		T07	&	22.94 	&	22.37 	&	22.35 	&	23.46 	&	24.06 	\\  \hline
		T05	&	19.61 	&	27.52 	&	26.67 	&	19.73 	&	29.87 	\\  \hline
		T09	&	24.53 	&	29.87 	&	27.42 	&	24.56 	&	25.06 	\\  \hline
		T11	&	14.00 	&	21.30 	&	24.79 	&	62.13 	&	68.53 	\\  \hline
		T10	&	24.15 	&	24.02 	&	24.04 	&	23.99 	&	24.54 	\\  \hline
		B01	&	25.59 	&	30.64 	&	33.34 	&	26.71 	&	26.11 	\\  \hline
		B02	&	31.93 	&	27.99 	&	25.82 	&	31.97 	&	35.42 	\\  \hline
		Avg.	&	17.63 	&	22.95 	&	21.98 	&	22.19 	&	24.02 	\\  \hline

	\end{tabular}
\end{table}

As shown in Table \ref{FinalProgess_4PostProcessing}, the average EER of detection models  on the original   dataset is 17.63\%, which is much lower than that applied with audio post-processing operations.  
The performance of detection models   on utterances with 0.002 random Gaussian noise is better than that on utterances with 0.01 random Gaussian noise, expect for T11, T10 and B01. Specifically, the average EER of  detection models on NA-002 is 21.98\% which is lower than that 22.95\%  on  NA-01. It indicates  that the higher level Gaussian noise  added to  utterances,  the more difficult to be detected.
As shown in the last two columns of Table \ref{FinalProgess_4PostProcessing},  the average EERs of   detection models   on the MP3  and  AAC compression-decompression utterances are 22.19\% and  24.02\% respectively, which are much higher than that on   original   utterances. It is demonstrated that there is great performance degradation of detection methods  due to the lower bit-rate AAC compression-decompression operations.
In a word, it is shown that all four audio post-processing operations  degrade the performance of  synthetic  speech detection methods.

\section{Conclusion}

With the improvement of Mandarin TTS and VC technologies, the need for  an up-to-date Mandarin  dataset that can be used in  the research of synthetic speech detection also increases. In this paper, we introduce the FMFCC-A dataset  and provide a detailed analysis of two baseline methods and  top-performing submissions from the FMFCC-A. The FMFCC-A dataset  contains  10,000 genuine utterances  collected collected  from 58 speakers  and  40,000 synthesized utterances    generated by 11  Mandarin TTS systems and two Mandarin VC systems. 
We divide  the FMFCC-A dataset as  the training,  development and evaluation sets, and conduct four kinds of audio post-processing operations on valuation dataset.
The training and  development  datasets are used for training and optimizing the synthetic speech detection models, while  the evaluation dataset   focus  upon  evaluating the generalization of synthetic speech detection methods   to previously unknown speech synthesis systems and audio post-processing operations.
Through the detailed analysis of   two baseline methods and
the top-performing submissions from    the FMFCC-A, we demonstrate the  usefulness and challenge of FMFCC-A dataset.
At the same time, we also observe that previously unknown  speech synthesis systems  and  audio post-processing operations would significantly degrade the performance of synthetic  speech detection.

There are two main areas of future work regarding the FMFCC-A dataset.  
Although the FMFCC-A dataset is already a good source   for 
the research of detection of synthesized  Mandarin speech, it can be improved by aggregating more  genuine and synthetic utterances in Mandarin. In addition, more realistic audio post-processing operations will be introduced  to enhance the practical application of FMFCC-A dataset.  
With the FMFCC-A dataset created and published, we hope that the research community can use it to improve the detection of synthesized  Mandarin speech.



\bibliographystyle{splncs03}
\bibliography{mybibfile_DataSet}

\end{document}